\documentclass[twocolumn,showpacs,preprintnumbers]{revtex4}
\begin{document}

\title{Heat Dissipation from  Brownian Particles under Hydrodynamic Interactions}
\author{Kyung Hyuk Kim}
\email{kkim@u.washington.edu}
\affiliation{Department of Physics, University of Washington, Seattle, WA 98195}%Lines break automatically or can be forced with \\

\date{\today}% It is always \today, today,
             %  but any date may be explicitly specified

\begin{abstract}
%We extend our mesoscoipc themodynamics of Brownian single macromolecules into the case of multiplicative noise including hydrodynamic interaction, where Ito-Stratonovich dilemma exists.  We provide 
We study the non-equilibrium thermodynamics of  single Brownian macromolecules immersed in water solvent. They are under both a hydrodynamic interaction and a feedback control on their movement by an external agent.  The macromolecules are described by a Langevin equation with a multiplicative noise.  Work done by the macromolecules on the water solvent is dissipated as heat. Thus, the heat is expressed as the integration of an interacting force between the macromolecules and the water solvent along the position space trajectories of the macromolecules.  This integration is stochastic due to the Brownian motion of the macromolecules.  We show that the Stratonovich prescription of the integration is the unique physical choice. We also show that thermodynamic quantities such as heat, work, and entropy production, are derived without any ambiguity if both a diffusion matrix and external feedback control are known as priori.
\end{abstract}
\maketitle
%by providing heat dissipation from the macromolecules to their surrounding heat bath, which enables prescription-independent mesoscopic thermodynamics.
%The relation shows transverse frictional force as in vortex dynamics in homogeneous superconductors. Furthermore, detailed balance shows the transverse friction is symmetric under time reversal operation. 
% Valid PACS numbers may be entered using the \verb+\pacs{#1}+ command.

\newcommand{\mb}[1]{\mathbf{#1}}
\newcommand{\mbh}[1]{\mathbf{\hat{#1}}}
\newcommand{\Gammah}{\hat{\Gamma}}

\section{Introduction}
Brownian dynamics has been widely applied to nano-scale (macromolecular) driven systems such as AFM/DFM cantilevers \cite{Liang, Tamayo, Kim03}, motor proteins \cite{Juli97}, ion channels \cite{Colqu}, and tribology \cite{Urbakh04}.  These systems are mainly in non-equilibrium driven by external agents.  The design of efficient nano-scale systems requires mesoscopic non-equilibrium thermodynamics.   
 We have introduced mesoscopic heat, work, and entropy \cite{Kim03, Kim06, Seki97, Seifert05}.  We have provided a  rigorous thermodynamic analysis on a molecular refrigerator composed of both a Brownian harmonic oscillator and an external control agent who actively reduces the thermal fluctuation of the oscillator \cite{Kim03, Kim06}.  We have assumed that the interaction between the refrigerator and its surrounding is expressed as linear friction and Gaussian white noise.  

In this manuscript, we extend our thermodynamic analysis to study Brownian particles under \emph{multiplicative} noise.  When polymers are in solvent, they are under hydrodynamic interactions \cite{Doi, Hooger92}. Their Brownian motions are well described by a state-dependent diffusion process, i.e.,  multiplicative noise.   
Such noise appears in other diverse fields \cite{Sancho82}, e.g., motor proteins \cite{Bier96} and regulation of gene expression \cite{Hasty00}.  
Most of numerical and analytical analysis on the systems under the multiplicative noise have been focused on their stochastic dynamics rather than on their thermodynamics.  To our knowledge, the quantitative thermodynamic analysis is absent due to the following two reasons.

The first reason is the lack of physical concept in mesoscopic heat dissipated from a Brownian particle.  The mesoscopic heat is the mesoscopic work done by the Brownian particle on the solvent, so it is expressed as an integration of an interacting force between the particle and the solvent along  its position space trajectory of the particle: $Q(t) = -\int_{s=0}^{s=t} d\mb{x}(s) \cdot \mb{F}_{PS}(\mb{x}(s), \mb{v}(s))$ with $\mb{x}(s)$ ($\mb{v}(s)$) is the position (velocity) of the particle at time $s$ and $\mb{F}_{PS}$ is the force done on the particle by the solvent molecules and $t$ is time.  We have adopted without rigorous justification that the above integration along the trajectory is done with the Stratonovich prescription \cite{Qian-time, Kim03, Kim06}.  In this manuscript, we will show its justification.

The second reason is that the mesoscopic heat needs another prescription in $\mb{F}_{PS}(\mb{x}(s), \mb{v}(s))$.  The phase space trajectory of the particle is different for different prescriptions of the stochastic integration involved in $\mb{v}(t) = \int_0^t ds \mb{F}_{T}(\mb{x}(s), \mb{v}(s))$ with unit mass and with the total force on the particle $\mb{F}_{T}$.  This means that $\mb{F}_{PS}(\mb{x}(s), \mb{v}(s))$ is also dependent on the prescription in $\mb{v}$.  However, the dependence on the prescription is shown to be removed under a generalized Einstein relation \cite{Arnold00}.  Then, can one express heat in a form independent to the prescription in $\mb{v}$  under the relation?  We provide such expression of heat from energy balance.

We also answer a fundamental question, ``Does detailed balance guarantee equilibrium steady state?"    Recently, in Brownian systems without any multiplicative noise, the detailed balance has been shown to guarantee the systems to reach at equilibrium steady state \cite{Qian-time, Kim03}.   However, as presented here, in systems under multiplicative noise, this is not true.

%The detailed balance stems from microscopic reversibility in Hamiltonian governing the microscopic equations of motion \cite{Mazur, Kampen}. In its derivation, an ergodic hypothesis is used, which is assumed to hold in our Brownian dynamics????.  Since a closed system of many classical particles, comprising water molecules and single macromolecules, satisfies the microscopic reversibility,  the detailed balance is expected to hold in mesoscopic description of the macromolecules, i.e., the detailed balance is a necessary condition on equilibrium.  

This paper is organized as follows: from Sec.\ref{model} to Sec.\ref{sec-detailed-ep}, we study a model of macromolecules in \emph{closed} heat bath, where the system reaches at an equilibrium steady state.  In Sec.\ref{model}, a model of the macromolecules is introduced using both a Langevin equation and its corresponding Fokker-Planck equation.  Mesoscopic heat dissipated from Brownian particles is introduced from mesoscopic energy balance with Stratonovich prescription of stochastic integration involved in the integration of force along the past spatial trajectories.  In Sec.\ref{sec-Einstein}, the generalized Einstein relation is derived to guarantee a Maxwell-Boltzmann distribution in an equilibrium steady state.  We show that the relation makes the mesoscopic heat independent to the prescription of stochastic integration of the Langevin equation.  In Sec.\ref{sec-entropy}, using the definite form of heat dissipation, we derive entropy balance and show the H-function, i.e., free energy, becomes maximized in equilibrium state.  In Sec.\ref{sec-detailed} and \ref{sec-detailed-ep}, we investigate the relationship among detailed balance, an equilibrium state, and the H-function.  In Sec.\ref{sec-nonequil}, we consider a \emph{driven} system by feedback controls and reformulate all the previous results for a non-equilibrium steady state. Finally, in Sec.\ref{sec-proof}, we prove that the Stratonovich prescription of stochastic integration involved in the definition of mesoscopic heat is the unique physical choice.

\section{A Model of Single Macromolecules under Multiplicative Noise} \label{model}
We consider a \emph{closed} system composed of both a macromolecule and its surrounding isothermal water solvent.  Following the general theory of polymer dynamics \cite{Doi}, the macromolecule itself is described by a bead spring model with Hamiltonian $H = \Sigma_\alpha \frac{\mb{p}_\alpha^2}{2m_\alpha} + U_{int}(\mb{x}_1, \cdots, \mb{x}_N)$ where  $\mb{x}_\alpha$ and $\mb{p}_\alpha$ are the 3-D coordinate- and momentum-vectors of the $\alpha$-th bead of the macromolecule, respectively, and $U_{int}$ is an internal potential of the macromolecule. The random collisions between water molecules and the beads are modeled by a multiplicative noise and frictional force $\mathbf{f}(\mb{v}_1 ,\cdots ,\mb{v}_N  )$ with $\mb{v}_i$ the velocity of the $i$-th bead. This is because the bead is assumed to be much larger than water molecules in heat bath and thus the time scales of the two can be separated \cite{Shea96, Shea98}. For simplicity but without losing any generality, we consider a macromolecule simply as one point-like bead. Its internal energy $H$ becomes $H= v^2/2$ with a unit mass.   The dynamics of the bead can be described by the following Langevin equation, 
\begin{equation}
\frac{d\mb{v}}{dt} = \mathbf{f}(\mb{v}) + \mbh{\Gamma}(\mb{v}) \cdot \mb{\xi}(t),
\label{nonlinear-langevin}
\end{equation}
where $\xi$ is Gaussian white noise satisfying $\langle \mb{\xi}(t)_i \mb{\xi}(t^\prime)_j \rangle = \delta_{ij}\delta(t-t^\prime)$ with $i=x,y,z$ in 3-D.  Note that the noise term has a state-dependent amplitude, $\mbh{\Gamma}(\mb{v})$, and such state-dependent noise, $\mbh{\Gamma}(\mb{v})\cdot \mb{\xi}$, is called a multiplicative noise, which incorporates hydrodynamic interactions.  For example, spherical hard particles with finite radius $R$ undergoes hydrodynamic interactions of nonlinear fluctuation force in Oseen approximation \cite{Hermans},  
\begin{equation}
(\mbh{\Gamma} \mbh{ \Gamma}^T)_{ij} = T \zeta \{ (1+\frac{9}{16} \frac{\rho R}{\eta} v) \delta_{ij} - \frac{3}{16}\frac{\rho a}{\eta} \frac{v_i v_j}{v}   \},
\label{oseen}
\end{equation}
where $\zeta$ is the frictional coefficient due to the interaction between solvent and the particle, and $\rho$ is the density of solvent, and $v$ the modulus of $\mb{v}$.  The frictional force, $\mb{f}(\mb{v})$, is shown later to be given by Eq.(\ref{fprime}) (\ref{fric2})and (\ref{f-hydro}) if $\mbh{\Gamma}$ is known. 
 We will discuss the above example more in detail in Sec. \ref{sec-Einstein}.

We should note that Eq.(\ref{nonlinear-langevin}) is meaningless without integration prescription since a pulse in white noise causes a finite jump in $\mb{v}$ and then the value of $\mb{v}$ in $\mbh{\Gamma}(\mb{v})$ needs to be prescribed.  Two popular prescriptions by Ito and Stratonovich have been widely used.  The Ito prescription takes the value of $\mb{v}$ in $\mbh{\Gamma}(\mb{v})$ before the jump in $\mb{v}$ caused by the white noise and the Stratonovich prescription takes it as the middle point value of $\mb{v}$ before and after the jump.     A Langevin equation with each prescriptions can be converted into Langevin equations with the other prescriptions \cite{Gardi}. For the ease of calculation, we convert Eq.(\ref{nonlinear-langevin}) into a corresponding Ito-prescribed form, 
\begin{equation} 
\frac{d\mb{v}}{dt} = \mb{f^\prime}+  \mb{\hat{\Gamma}}(\mb{v}) \cdot \mb{\xi}(t),
\label{Ito}
\end{equation}
where 
\begin{equation}
f^\prime_i(\mb{v}) \equiv f_i(\mb{v}) + a \hat{\Gamma}_{kj}(\mb{v}) \partial_j \hat{\Gamma}_{ki}(\mb{v}),
\label{fprime}
\end{equation}
with $a=0 ~ (1/2)$ for Ito (Stratonovich)-prescribed Eq.(\ref{nonlinear-langevin}). Note that the Einstein summation rule is used.  

Let's consider energy balance.  The change of mechanical energy of the macromolecule, $dH(X_t, Y_t)$, is the same as the work done on the macromolecule by all the external forces, i.e., $dH(X_t, Y_t) = (f^\prime+\hat{\Gamma} \cdot \xi) \circ dX$, where $\circ$ indicates that the stochastic integration is done in Stratonovich way.  We will be shown in Sec.\ref{sec-proof} that the  Stratonovich prescription for the definition of $dH$ is the unique physical choice.  Since the internal energy changes by heat dissipation and absorption by the bead through interaction with the surrounding heat bath, we identify \cite{Seki97,Kim03} 
\begin{equation}
dQ(X_t,Y_t) \equiv -dH(X_t, Y_t)=-(f^\prime+\hat{\Gamma} \cdot \xi) \circ dX.
\label{defheat}
\end{equation} 
This indicates how much heat is dissipated (absorbed) to (from) the surrounding water heat bath from (to) the bead located at $(X_t,Y_t)$ at time $t$ during time interval $dt$ for a stochastic process.   Using Eq.(\ref{Ito}), we derive 
\begin{equation}
\frac{dQ}{dt}  = H_d + v_i \cdot \Gamma_{ij} \cdot \xi_j,  \label{dQ}
\label{stochasticheat}
\end{equation}
where $H_d \equiv -\mb{v} \cdot  \mb{f}^\prime  - \frac{1}{2} Tr[\mbh{\Gamma} \mbh{\Gamma}^T]$, and Eq.(\ref{dQ}) is integrated with the Ito prescription.  For the detailed derivation of Eq.(\ref{dQ}), see \cite{Kim-prescription}.  

The ensemble average of heat dissipated up to time $t$ for a stochastic process, $Q(\{\mb{v}(s)\};\{0 \leq s \leq t \})$, is $ E[Q] = \int_0^t E[H_d(v(s))] ds.$  For a sufficiently small time interval $(t, t+\Delta t)$, the average amount of heat dissipated is $E[H_d]\Delta t$.  The heat dissipation rate ($h_d$) at time $t$ can then be defined as
\begin{equation}
h_d(t) \equiv E[H_d( \mb{v}(t))] = \int d\mb{v} \{ -\mb{v} \cdot  \mb{f}^\prime  - \frac{1}{2} Tr[\mbh{\Gamma} \mbh{\Gamma}^T] \} P(\mb{v},t),
\label{dissipation}
\end{equation}
where $P(\mb{v},t)$ is a probability distribution function satisfying a Fokker-Planck equation corresponding to Eq.(\ref{Ito}),
\begin{equation}
\frac{\partial P}{\partial t} = \frac{1}{2} \partial_i \partial_j \{(\hat{\Gamma} \hat{\Gamma}^T)_{ij} P\} - \partial_i (f^\prime_iP) = \mathcal{L}P.
\label{fokker}
\end{equation}

Eq.(\ref{dissipation}) implies fluctuation dissipation relation ($h_d(t=\infty)=0$) in equilibrium as shown in Sec.\ref{sec-Einstein}: frictional dissipation from the bead to the heat bath is balanced by heat absorption  by fluctuation from the heat bath to the bead.  In addition, the formula of Eq.(\ref{dissipation}) is shown in Sec.\ref{sec-nonequil} not to be changed  in a driven system  and  to imply the breakdown of the fluctuation dissipation relation.

\section{Generalized Einstein Relation and Heat Dissipation Rate} \label{sec-Einstein}
In this section, generalized Einstein relation is proposed for the correct equilibrium steady state distribution of the Fokker-Planck equation Eq.(\ref{fokker}), i.e., Maxwell-Boltzmann distribution.  This relation is shown to make the mesoscopic heat, Eq.(\ref{stochasticheat}), independent to stochastic integration prescription of the Langevin equation, Eq.(\ref{nonlinear-langevin}), and vanish at equilibrium steady state.  In addition, this relation shows interesting information on the form of frictional force which includes (1) nonlinear frictional force stemming from hydrodynamic interactions, e.g., dissipative force in dissipative particle dynamics \cite{Hooger92} and frictional force in a Zimm model \cite{Doi} and (2) transverse force in vortex dynamics of homogeneous superconductors \cite{Ao99}.

By substituting Maxwell-Boltzmann distribution function ($P_e(\mb{v}) \equiv C e^{-v^2/T}$ with $C$ normalization constant and with $k_B=1$ unit) into Eq.(\ref{fokker}), we can find the form of frictional force $\mb{f}(\mb{v})$.
The r.h.s. of Eq.(\ref{fokker}) is simplified as
\begin{eqnarray*} 
\lefteqn{\frac{1}{2} \partial_i \partial_j \{(\Gammah\Gammah^T)_{ij} P_{e}\}- \partial_{i}( f^\prime_i P_{e} )} \\
		&=& \partial_i \Bigl[ \frac{1}{2}\partial_j \{(\Gammah \Gammah^T)_{ij} P_e \} - f^\prime_i P_e    \Bigr]\\
		&=& \partial_i \Bigl[ \{\frac{1}{2}\partial_j (\Gammah \Gammah^T)_{ij} -\frac{1}{2}(\Gammah \Gammah^T)_{ij}\frac{v_j}{T}  - f^\prime_i \} P_e\Bigr].
\end{eqnarray*}
Therefore, 
\begin{equation}
f^\prime_i(\mb{v}) = \frac{1}{2} \partial_j(\Gammah(\mb{v}) \Gammah^T (\mb{v}))_{ij} - \frac{1}{2}(\Gammah(\mb{v}) \Gammah^T(\mb{v}))_{ij}\frac{v_j}{T} + b_i(\mb{v}),
\label{fric2}
\end{equation}
where $\mb{b}(\mb{v})$ is an arbitrary solution of $\partial_i (b_i(\mb{v})P_e(\mb{v})) = 0$. 
%From Eq.(\ref{fprime}), $\mb{f}(\mb{v})$ becomes 
%\begin{eqnarray}
%f_i(\mb{v}) &=& \frac{1}{2} \partial_j(\Gammah(\mb{v}) \Gammah^T (\mb{v}))_{ij} -a \Gammah_{kj}(\mb{v}) \partial_j %\Gammah_{ki}(\mb{v}) \nonumber \\
%	&&- \frac{1}{2}(\Gammah(\mb{v}) \Gammah^T(\mb{v}))_{ij}\frac{v_j}{T} + b_i(\mb{v}).
%\label{fric3}
%\end{eqnarray}

Now, by substituting Eq.(\ref{fric2}) to Eq.(\ref{dissipation}), heat dissipation rate $h_d(t)$ becomes 
\begin{eqnarray}
h_d(t)&=& \int d\mb{v} \Bigl[ \{ - \frac{1}{2}\partial_j (\Gammah \Gammah^T)_{ij}+\frac{1}{2}(\Gammah \Gammah^T)_{ij}\frac{v_j}{T}-b_i\} v_i \nonumber \\
	&&- \frac{1}{2}Tr[\Gammah \Gammah^T] \Bigr] P(t). \label{3Dhd} 
\end{eqnarray}
In equilibrium, by substituting $P_e(\mb{v})$ to the above equation,
\begin{eqnarray}
h_d(\infty)	&=& \int d\mb{v} \Bigl[ \frac{1}{2}(\Gammah \Gammah^T)_{ij}\partial_j(v_i P_e) \nonumber \\
	&& + \{ \frac{1}{2}(\Gammah \Gammah^T)_{ij}\frac{v_i v_j}{T}- \frac{1}{2}Tr[\Gammah \Gammah^T] \} P_e - b_i v_i P_e \Bigr] \nonumber \\
	&=&-\langle \mb{b} \cdot \mb{v} \rangle_{e}, \label{gel-Ein}
\end{eqnarray}
where $\langle \cdot  \rangle_{e}$ means average over equilibrium distribution function $P_e(\mb{v})$.
The heat dissipation rate is expected to vanish in equilibrium, so we propose a generalized Einstein relation:
\begin{equation}
\mbox{Eq.(\ref{fric2})}\quad \mbox{and}   \quad \langle \mb{b} \cdot \mb{v} \rangle_{e}=0,
\label{gen-einstein-rel}
\end{equation}
where $\mb{b}$ an arbitrary solution of $\partial_i (b_i(\mb{v})P_e(\mb{v})) = 0$.  Since all the above procedure can be reversed back, the generalized Einstein relation also guarantees Maxwell-Boltzmann distribution for closed systems.  Furthermore, the relation lets heat dissipation vanish,  entropy production vanish, and an average flux $\langle \mb{v} \rangle_e$ trivially vanish as shown in Sec.\ref{sec-entropy}.  

Under the generalized Einstein relation, the Fokker-Planck equation Eq.(\ref{fokker})  becomes independent to the choice of integration prescription of the Langevin equation, i.e., the constant, $a$ \cite{Arnold00}. The definition of the mesoscopic heat Eq.(\ref{defheat}) derived from the energy balance makes Eq.(\ref{stochasticheat}) $a$-independent  since $f^\prime$ becomes $a$-independent from the generalized Einstein relation (See Eq.(\ref{fric2})).  This makes  thermodynamic quantities such as entropy production, work and internal energy $a$-independent.   We conclude that \emph{when a Langevin equation with multiplicative noise is given, we can construct mesoscopic thermodynamics independent to the prescription of the stochastic integration in the Langevin equation as long as its steady state follows Maxwell-Boltzmann distribution.}   In Sec.\ref{sec-nonequil}, we consider a non-equilibrium system where the generalized Einstein relation still holds.  Such system will be shown later for both the Fokker-Planck equation and the mesoscopic heat, Eq.(\ref{stochasticheat}), still to be independent to the stochastic integration prescription of its Langevin equation.  We reach at a more general conclusion that \emph{as long as the generalized Einstein relation holds, one can construct mesoscopic non-equilibrium thermodynamics independent to the prescription of the stochastic integration in the Langevin equation.}  The validity of the generalized Einstein relation  and the proposed heat must be tested by measuring thermodynamic quantities using molecular dynamics simulations. 

Note also that since Fokker-Planck equation Eq.(\ref{fokker}) and the mesoscopic heat equation Eq.(\ref{stochasticheat}) are expressed in terms of diffusion coefficient $\mbh{\Gamma}\mbh{\Gamma}^T$, \emph{one can numerically or analytically predict all the thermodynamic quantities by measuring the diffusion coefficient}.  For example, polymers in solvent under hydrodynamic interactions (See Eq.(\ref{oseen})) undergoes  frictional force $f^\prime$,
\begin{equation}
f_i^\prime= - \frac{\zeta}{2} v_i (1+\frac{3}{8} \frac{\rho R}{\eta} v),\label{f-hydro}
\end{equation}
 and heat is dissipated on average by
\begin{eqnarray}
h_d(t) &=& \int d\mb{v} \Big[  \zeta ( \frac{v^2}{2} - \frac{3T}{2} ) \nonumber \\
	&& + \frac{3}{4} \frac{\rho R }{\eta} \zeta v (-T + \frac{v^2}{4})  \Big] P(\mb{v},t). \label{hd-hydro}
\end{eqnarray}
The heat dissipation rate, as expected,  vanishes at equilibrium; Substituting $P_e$ into Eq.(\ref{hd-hydro}), the first term in $h_d$ vanishes from equi-partition theorem and the second term vanishes since the order of $v$ is odd.   Eq.(\ref{f-hydro}) and (\ref{hd-hydro}) become useful in a driven system, where one can analytically predict average heat dissipation rate in steady state once the steady state distribution is known, or if not known, at least  one can get numerics by Eq.(\ref{dQ}).

The frictional force term $b(\mb{v})$ can be transverse force to velocity $\mb{v}$.  One of the concrete example is vortex dynamics in homogeneous superconductors \cite{Ao99}.  This vortex system is a closed system since a transverse force $b(\mb{v})$ on vortex is applied by magnetic field produced by superfluid circulation.

If a macromolecule is immersed in a general isotropic frictional medium \cite{Klimon90} such as  the frictional force $\mb{f}^\prime(\mb{v})$ is expressed as $\mb{f}^\prime(\mb{v}) = - \gamma(v) \mb{v}$ and the fluctuation force is as $\mb{\Gammah}_{ij}(\mb{v}) = \delta_{ij}\Gamma(v)$, the frictional force $\mb{f}^\prime$ becomes
\begin{equation}
f^\prime_i(\mb{v}) = ( -\frac{\Gamma(v)^2}{2T} +\frac{1}{2 v} \frac{\partial \Gamma(v)^2}{\partial v} ) v_i,
\end{equation}
where we used the fact that $\mb{b}$ vanishes from one of the generalized Einstein relation $\langle \mb{b} \cdot \mb{v} \rangle_{e}=0$ since $\mb{b}$ cannot be parallel to $\mb{v}$.

\section{entropy production rate} \label{sec-entropy}
In a non-equilibrium system, entropy is produced (created) from its inside.  For example, let two heat baths having different temperature connected together.  Then, the total entropy change of the two heat bath is $dS = dQ_{1 \rightarrow 2} (1/T_2 - 1/T_1)>0$, where $dQ_{1 \rightarrow 2}$ is the heat transferred from heat bath 1 to heat bath 2 and $T_{1(2)}$ is the temperature of heat bath 1(2).  In equilibrium, entropy production vanishes.  In macromolecular system immersed in solvent, the macromolecule is  not in a quasi-static process, while the heat bath is because the macromolecule has a small number of degrees of freedom; there is no boundary layer that the fluctuation caused by interaction with heat bath disappears.   The entropy change of heat bath, $dS_H$, is $dQ_{M \rightarrow H} / T_H$, where $dQ_{M \rightarrow H}$ is the heat transferred from the macromolecule to the heat bath and $T_H$ is the temperature of heat bath.  However, the entropy change in the macromolecule, $dS_M$, is not $dQ_{H \rightarrow M}/T_M$, where $T_M$ is the temperature of the macromolecule if it can be defined.  How can one construct $dS_M$?

We consider Gibbs entropy, $S_M(t) \equiv - \int d\mb{v} P(\mb{v}) \ln P(\mb{v})$ \cite{Schnak76, Jou99} and $T \equiv T_H$.   
It will be shown in this section that entropy balance is expressed as,
\begin{equation}
\frac{d(S_M(t)+S_H(t))}{dt}=e_{p}(t) \geq 0,
\label{entropy-balance}
\end{equation}
where 
\begin{equation}
e_p(t) \equiv \frac{1}{T}\int \Pi_i(t) J_i(t) d\mb{v},
\label{ep}
\end{equation}
and $\Pi$ is a thermodynamic force defined as the sum of the second term of the frictional force expressed in Eq.(\ref{fric2}) and Onsager's thermodynamic force:
\[
\Pi_i(t) \equiv - (1/2T)(\Gamma \Gamma^T)_{ij}(v_j + T \partial_j \ln P(t))
\]
and $J(x,y,t)$ is a thermodynamic flux corresponding to the thermodynamic force $\Pi$ and is  defined as:
\[
J_i(t) \equiv -(v_i + T \partial_i \ln P(t)) P(t).
\]
$J_i$ is  the sum of both the velocity of the macromolecule and the diffusion flow in momentum space.  Note that, as in macroscopic non-equilibrium thermodynamics \cite{Mazur}, $e_p(t)$ is expressed as a product of thermodynamic force and its corresponding flux.  Note also that $e_p(t)$ is always non-negative. This implies the 2nd law of thermodynamics. In the equilibrium steady state, the entropy production vanishes: the macromolecule can be considered to be in quasi-static process and $T_M$ well defined to be $T$.

Let's start the derivation of Eq.(\ref{entropy-balance}).  $T$ denotes heat bath temperature. For the ease of calculation, we introduce $\Xi \equiv (1/2T)\Gamma \Gamma^T$. 
\begin{eqnarray}
\frac{dS_M}{dt} &=& -\frac{d}{dt} \int P \ln P d\mb{v} \nonumber  = - \int \frac{\partial P}{\partial t} \ln P d\mb{v} \nonumber \\ 
	&=&-\int \partial_i(-f_i^\prime P + T \partial_j( \Xi)_{ij} P + T \Xi_{ij} \partial_j P ) \nonumber \\
		&&\times \ln P d\mb{v} \nonumber \\
	%&=&\int (-f_i^\prime P + \partial_j(T \Xi)_{ij} P + T \Xi_{ij} \partial_j P ) \partial_i \ln P dv \nonumber \\
	&=&\int (-f_i^\prime  + T\partial_j( \Xi)_{ij}  + T \Xi_{ij} \partial_j \ln P )  \partial_i P  d\mb{v} \nonumber \\
	%&=&\int (-f_i^\prime  + b_i+ \partial_j(T \Xi)_{ij}  + T \Xi_{ij} \partial_j \ln P )  \partial_i P  dv - \int b_i \partial_i P dv\nonumber \\
	&=& \int \Xi_{ij}(v_j + T \partial_j \ln P ) \partial_i P dv - \int b_i \partial_i P d\mb{v} \nonumber \\
	&=& \frac{1}{T}\int \Xi_{ij}(v_j + T \partial_j \ln P)(v_i + T \partial_i \ln P) P d\mb{v} \nonumber \\
	&&- \frac{1}{T}\int \Xi_{ij}(v_j+ T \partial_j \ln P) v_i P d\mb{v} \nonumber \\
	&&-\int b_i \partial_i P d\mb{v} \label{entropy-bal2},
\end{eqnarray}
where   the first term is entropy production rate $e_p(t)$ and the second and the third are the entropy change in heat bath due to heat dissipation from the macromolecule,
\begin{equation}
\frac{dS_H(t)}{dt} \equiv \frac{1}{T}\int J_i(t) (-\Xi_{ij}v_i) d\mb{v} + \int b_i \partial_i P d\mb{v}.
\label{hd}
\end{equation}
$dS_H(t)/dt$ can be simplified as
\begin{eqnarray*}
T\frac{dS_H(t)}{dt}&=	& \int -(f_i^\prime - \partial_j(T \Xi_{ij}) - b_i - T \Xi_{ij}\partial_j \ln P) v_i P d\mb{v} \nonumber \\
	&&- T\langle \partial_i b_i \rangle\\
	%&=& \int -f_i^\prime v_i P dv + \int \{ \partial_j (T \Xi_{ij})P + T \Xi_{ij} \partial_j P  \} v_i d\mb{v} + \langle \mb{b}\cdot \mb{v}\rangle - T\langle \partial_i b_i \rangle\\
	%&=& \int -f_i^\prime v_i P dv + \int \partial_j ( T \Xi_{ij}P) v_i dv + \langle \mb{b}\cdot \mb{v}\rangle - T\langle \partial_i b_i \rangle\\
	&=& \int (-f_i^\prime v_i - Tr[T \Xi_{ij}])P d\mb{v} + \langle \mb{b}\cdot \mb{v}\rangle- T\langle \partial_i b_i \rangle\\
	&=& h_d(t) + \langle v_i b_i -T\partial_i b_i \rangle.
\end{eqnarray*}
Now, we have an unexpected extra term and this term, however, is proved to vanish: $\mb{b}$ is an arbitrary solution of  $\partial_i (b_i P_e) =0$, which  is equivalent to $T\partial_i b_i - v_i b_i=0$.  Therefore, 
\begin{equation}
\frac{dS_H(t)}{dt} = \frac{h_d(t)}{T}. \label{dsh}
\end{equation}
This confirms that the heat bath is in quasi-static process.

\section{detailed balance and potential conditions}\label{sec-detailed}
A system is called to be \emph{microscopically reversible} \cite{Tolman, Mazur, Kampen} when the microscopic equations of motion governed by a Hamiltonian is invariant under time reversal operation.  From this microscopic reversibility,  detailed balance is derived with ergodic hypothesis \cite{Mazur, Kampen}.  We are interested in a closed system of many classical particles, comprising solvent molecules and macromolecules, of which the equation of motion satisfies the microscopic reversibility.  Therefore, the detailed balance is expected to hold in mesoscopic description of the Langevin equation Eq.(\ref{Ito}).  

In Brownian systems without any multiplicative noise, the detailed balance has been shown to be a \emph{sufficient} condition for equilibrium steady state \cite{Qian-time, Kim03}.  However, as rooted from the microscopic reversibility, the detailed balance will be shown in this section to be just a necessary condition for equilibrium steady state. 
%The detailed balance gives information on the properties of constant coefficients related to time reveral operation.  For example, magnetic field is asymmetric under time reversal operation since  the field is produced by the linear sum of the velocities of charged particles that are asymmetric under time reversal operation.  Another example is overdamped case of  a macromolecule immersed in solvent.    MORE DESCRIPTION NEEDED

Now, let's derive well-known \emph{potential conditions} \cite{Graham71} in discrete time and continuous space, which can be trivially extended to the case of continuous time and continuous space.  The time increment is denoted by $\epsilon$ and $t_m \equiv m \epsilon$ with $m$ an positive integer. When the system is Markovian, we can introduce transfer matrix, $\hat{T}(\{C_i \})$, satisfying $|P\rangle_{t_{m+1}} = \hat{T}(\{ C_i \})|P\rangle_{t_{m}}$.  
The detailed balance \cite{Kampen} is expressed as,
\begin{eqnarray}
\lefteqn{\langle \mb{v}_{m+1} | \hat{T}(\{C_i \})|\mb{v}_m \rangle \langle \mb{v}_m | P;\{C_i\}\rangle_{e}} \nonumber \\
 &=\langle -\mb{v}_m | \hat{T}(\{\epsilon_i C_i \})|-\mb{v}_{m+1} \rangle \langle -\mb{v}_{m+1} | P;\{\epsilon_i C_i\} \rangle_{e},
\label{ext-detail0}
\end{eqnarray}
where $\epsilon_i$ is $1$ for a constant coefficient, $C_i$, if it is symmetric under time reversal operation,  $-1$ if anti-symmetric, and other value if neither symmetric nor anti-symmetric determined from microscopic origin \cite{Kim03, Kampen}.  $\mb{v}_i$ denotes $\mb{v}$ at time $t_i$ along a process by transfer matrix $\hat{T}(\{ C_i \})$.  By integrating out $\mb{v}_m$ in Eq.(\ref{ext-detail0}), we get $\langle \mb{v} | P\rangle_e = \langle -\mb{v} | \tilde{P}\rangle_e$, where $|P\rangle_e \equiv | P;\{C_i\} \rangle_e$, $|\tilde{P}\rangle_e \equiv | P;\{\epsilon_i C_i\} \rangle_e$.   Therefore, Eq.(\ref{ext-detail0}) becomes
\begin{eqnarray}
\lefteqn{\langle \mb{v}_{m+1} | \hat{T}(\{C_i \})|\mb{v}_m \rangle \langle \mb{v}_m |P\rangle_e } \nonumber \\
 & = \langle -\mb{v}_m | \hat{T}(\{\epsilon_i C_i \})|-\mb{v}_{m+1} \rangle \langle \mb{v}_{m+1} |{P}\rangle_e.
\label{ext-detail}
\end{eqnarray}
The transfer matrix is related to the linear operator $\hat{\mathcal{L}}$ of Fokker-Planck equation Eq.(\ref{fokker}) as $\hat{T}= I + \epsilon \hat{\mathcal{L}}$. Now, Eq.(\ref{ext-detail}) is re-expressed as 
\begin{equation}
\langle v|\hat{\mathcal{L}}^\dagger |v^\prime \rangle \langle v|P \rangle_e = \langle -v|\hat{\widetilde{\mathcal{L}}}|-v^\prime \rangle \langle v^\prime |P\rangle_e,
\label{d2}
\end{equation}
where $\hat{T}(\{\epsilon_i C_i \})\equiv I+\epsilon \hat{\widetilde{\mathcal{L}}}$ is used.
Eq.(\ref{d2}) becomes 
\begin{eqnarray}
\mathcal{L}^\dagger_v \delta(v-v^\prime) P_e(v^\prime) = \frac{1}{P_e(v)} \widetilde{\mathcal{L}}_{-v} \delta(v-v^\prime) {P}_e(v^\prime)^2 \nonumber \\
	\quad = \frac{1}{P_e(v)} \widetilde{\mathcal{L}}_{-v} {P}_e(v) \delta(v-v^\prime) {P}_e(v^\prime).
\end{eqnarray}
We find symmetry in the operator $\hat{\mathcal{L}}$, 
\begin{equation}
\mathcal{L}^\dagger_v  = \frac{1}{{P}_e(v)}\tilde{\mathcal{L}}_{-v} {P}_e(v). 
\label{rev1}
\end{equation}
From Eq.(\ref{fokker}), $\mathcal{L}^\dagger = \frac{1}{2} A_{ij} \partial_i \partial_j + f_i^\prime \partial_i$, where $\hat{A} \equiv \hat{\Gamma}\hat{\Gamma}^T$.  The r.h.s. of Eq.(\ref{rev1}) is expressed as
\begin{eqnarray*}
\lefteqn{\frac{1}{{P}_e(v)}\tilde{\mathcal{L}}_{-v} {P}_e(v)} \\ 
	&=&\frac{1}{P_e}\{ \frac{1}{2} \partial_i \partial_j \tilde{A}_{ij}(-v) P_e(v) - \partial_i \tilde{f}_i^\prime (-v) P_e(v) \}  \\
	&=& \frac{1}{2} \tilde{A}_{ij}(-v)\partial_i \partial_j  + \frac{1}{P_e}(\partial_i \tilde{A}_{ij}(-v) P_e) \partial_j  + \tilde{f}_i^\prime(-v) \partial_i \\
	 &&+ \frac{1}{P_e}(\frac{1}{2} \partial_i \partial_j \tilde{A}_{ij}(-v)P_e + \partial_i \tilde{f}^\prime_i (-v) P_e),
\end{eqnarray*}
where $\tilde{A}(v) \equiv A(v;\{\epsilon_i C_i\})$ and $\tilde{f}^\prime (v) \equiv f^\prime (v; \{\epsilon_i C_i \})$. 
Finally, by matching term by term, we derive a set of conditions well known as potential conditions \cite{Graham71}, 
\begin{eqnarray}
\partial_i \ln P_e(\mb{v})	&=&(\Gammah \Gammah^T)^{-1}_{ij} [ -f_j^\prime(-\mb{v};\{\epsilon_i C_i\}) + f_j^\prime(\mb{v}) \nonumber \\
	&& - \partial_k (\Gammah \Gammah^T)_{kj}] \label{rev3} 
\end{eqnarray}
\begin{eqnarray}
\lefteqn{\mbh{\Gamma}(\mb{v};\{C_i\})\mbh{\Gamma}^T(\mb{v};\{ C_i \})} \nonumber \\
&&	\quad \quad \quad  \quad \quad = \mbh{\Gamma}(-\mb{v};\{\epsilon_i C_i\})\mbh{\Gamma}^T (-\mb{v}; \{ \epsilon_i C_i \}).  \label{rev2}
\end{eqnarray}
The detailed balance Eq.(\ref{ext-detail}), the symmetry relation in operator $\mathcal{L}$ Eq.(\ref{rev1}), and the potential conditions Eq.(\ref{rev3}) and (\ref{rev2}) are all equivalent in Markovian systems with $P_e(\mb{v};\{C_i\}) = P_e(-\mb{v};\{\epsilon_i C_i\})$ \cite{Ito78, Graham71}.

%\subsection{1-D Brownian Molecule under Nonlinear Frictional Force}
%In equilibrium, detailed balance must hold and it applies additional conditions on the frictional force.   The potential condition, Eq.(\ref{rev3}) and (\ref{rev2}), becomes 
%\begin{eqnarray}
%\frac{d \ln P_{e}(v)}{dv}	&=&\frac{1}{\Gamma(v)^2 }( -f^\prime(-v;\{ \epsilon_i C_i \}) + f^\prime(v; \{C_i\}) +(a-1) \frac{d \Gamma(v)^2}{dv}) \label{pot1}\\
%\Gamma(v; \{C_i\})^2 &=& \Gamma(-v; \{\epsilon_i C_i\})^2. \label{pot2}
%\end{eqnarray}
%Eq.(\ref{pot2}) implies that $f(v)$ and $f^\prime(v)$ are antisymmetric under time reversal from Eq.(\ref{1Dfric}) and (\ref{1Dfprime}), and makes Eq.(\ref{pot1}) trivially hold.  If some terms of the frictional force are odd in velocity, their coefficients are symmetric under time reversal.  If even, then they are antisymmetric.  Note that since the coefficients come from integrating out microscopic information of the solvent molecules, the coefficients can be symmetric, antisymmetric, etc. \cite{Zwan01}  Also note that the detailed balance does not restrict further the form of the frictional force Eq.(\ref{1Dfric}).  

%Frictional force between two macroscopic objects are symmetric under time reversal since its contituting fundamental forces such as electro-magnetic force and gravitational force are symmetric under time reversal.  In mesoscopic system, the frictional force $f(v)$ is, however, antisymmetric!  I DON'T KNOW HOW TO DEAL WITH THIS ISSUE.

By substituting Eq.(\ref{fric2}) into Eq.(\ref{rev3}) and (\ref{rev2}), the potential condition becomes
\begin{equation}
\partial_i \ln P_e(\mb{v})	=-\frac{v_i}{T} + (\Gammah \Gammah^T)^{-1}_{ij} \{ -b_j(-\mb{v};\{\epsilon_i C_i\}) + b_j(\mb{v})\} \label{r1}
\end{equation}
\begin{equation}
\mbh{\Gamma}(\mb{v};\{C_i\})\mbh{\Gamma}^T(\mb{v};\{ C_i \}) = \mbh{\Gamma}(-\mb{v};\{\epsilon_i C_i\})\mbh{\Gamma}^T (-\mb{v}; \{ \epsilon_i C_i \}). \label{r2} 
\end{equation}
We can find the property of each terms of $\mb{f}^\prime$ under time reversal operation from Eq.(\ref{fric2}), (\ref{r1}), and (\ref{r2}): $\mb{b}(\mb{v})$ is symmetric and the rest of terms in Eq.(\ref{fric2}) are anti-symmetric.  If $\mb{b}(\mb{v})$ is odd (even) in velocity, then its coefficient must be anti-symmetric (symmetric) under time reversal, e.g, transverse force in vortex dynamics is odd in velocity and its coefficient, magnetic field, is anti-symmetric under time reversal operation \cite{Ao99}. For the rest of terms in Eq.(\ref{fric2}), however,  the coefficients of terms odd (even) in velocity are symmetric (anti-symmetric).  We can find time reversal properties of all the constant coefficients in $\mb{f}^\prime$  from the detailed balance condition  and, however, this condition does not apply any further restriction on the form of the frictional force.

\section{Detailed Balance v.s. $e_p=0$}\label{sec-detailed-ep}
In the region where the generalized Einstein relation is valid, if entropy production vanishes, equilibrium state reaches since $e_p = 0$ is equivalent to $ v_i + T \partial_i \ln P(t)=0$ from Eq.(\ref{ep}) and this equation guarantees $P(t)$ to be Maxwell distribution, $P_e$.  Therefore, $e_p=0$ is equivalent to equilibrium \cite{Kim03}.  Does detailed balance guarantee that system reaches in equilibrium?  No.  The detailed balance is necessary condition on equilibrium from Eq.(\ref{r1}) and (\ref{r2}) since time reversal property of transverse force, $\mb{b}(\mb{v})$, is not known as priori. Only when the transverse frictional force is symmetric under time reversal operation, the detailed balance guarantees equilibrium.  In summary, the relation among equilibrium, zero entropy production, and detailed balance is symbolically  expressed as
\begin{equation}
\mbox{Equilibrium} ~\equiv ~[e_p=0]~ \subset ~\mbox{Detailed ~Balance}.
\end{equation}

\section{A Driven system: Non-equilibrium Steady State} \label{sec-nonequil}
In the previous sections, we constructed mesoscopic thermodynamics for a closed system, which has an equilibrium steady state.  In this section, we extend all the previous analysis to a driven system, which has a non-equilibrium steady state.  We consider that macromolecules are under hydrodynamic interaction and are subject to a feedback control by an external agent.  As before, the hydrodynamic interaction is modeled by an Oseen tensor.  The feedback control is modeled by a non-conservative force, $\mb{g}(\mb{x})$ \cite{Kim-vdep}.  We assume that the driven system is near equilibrium in the sense that the generalized Einstein relation Eq.(\ref{gen-einstein-rel}) still holds.  Now, the non-conservative force, $\mb{g}(\mb{x})$, is added in the Langevin equation for a closed system, Eq.(\ref{nonlinear-langevin}):
\begin{equation}
\frac{d\mb{v}}{dt} = \mb{g}(\mb{x})+\mathbf{f}(\mb{v}) + \mbh{\Gamma}(\mb{v}) \cdot \mb{\xi}(t).
\label{langevin-noneuqil}
\end{equation}
The corresponding Langevin equation in Ito-prescribed form becomes
\[
\frac{d\mb{v}}{dt} = \mb{g}(\mb{x})+\mathbf{f^\prime}(\mb{v}) + \mbh{\Gamma}(\mb{v}) \cdot \mb{\xi}(t).
\]
Energy balance is expressed as $dH=dW-dQ$, where $dH$ is the change in internal energy that is work done by all external forces, i.e., $dH = [ \mb{g} + \mb{f^\prime} + \mbh{\Gamma}(\mb{v}) \cdot \mb{\xi} ]\circ d\mb{X}$ and $dW \equiv \mb{g}(\mb{X}) \circ d\mb{X}$ is work done by the control force $\mb{g}$.  The heat dissipation $dQ$ from the macromolecule to the surrounding heat bath becomes the same form as in a closed system: $dQ = -(f^\prime+\hat{\Gamma} \cdot \xi) \circ dX$.  
Fokker-Planck equation is changed to 
\[
\frac{\partial P}{\partial t} = \frac{1}{2} \partial_i \partial_j \{(\hat{\Gamma} \hat{\Gamma}^T)_{ij} P\} - \partial_i \{(g_i + f^\prime_i) P\}.
\]
Eq.(\ref{dQ}), (\ref{dissipation}), (\ref{fric2}), (\ref{entropy-balance}), (\ref{ep}) and (\ref{dsh}) are not altered. Its proof is just book-keeping of \cite{Kim-prescription} and Sec.\ref{sec-entropy}: entropy balance equation is not changed.  However, the steady state is not in equilibrium and the entropy production becomes positive, which means that the total entropy of both the macromolecule and heat bath constantly increases.  In other words, there is net positive heat dissipation from the macromolecule to the heat bath.

\section{The Definition of Heat with Stratonovich Prescription}\label{sec-proof}
As we discussed in Sec.{\ref{model}}, we have used Stratonovich prescription for the definition of $dQ$: $dQ \equiv -(\mb{f}^\prime + \mbh{\Gamma } \cdot \mb{\xi}) \circ d\mb{X}$.  It will be shown that the diffusion coefficient matrix, $\mbh{\Gamma}\mbh{\Gamma}^T$, becomes traceless, which is unphysical for diffusive systems, if we use other prescriptions in $dQ$.

Let $dQ \equiv -(\mb{f}^\prime + \mbh{\Gamma } \cdot \mb{\xi}) \bullet d\mb{X}$, where $\bullet$ indicates the stochastic integration is not in Stratonovich way. Eq.(\ref{dQ}) is changed to Ito-prescribed stochastic equation with multiplicative noise,
\[
\frac{dQ}{dt}  = H_d + v_i \cdot \Gamma_{ij} \cdot \xi_j
\]
where $H_d \equiv -\mb{v} \cdot  \mb{f}^\prime  - d Tr[\mbh{\Gamma} \mbh{\Gamma}^T]$. If $d=0$, it corresponds to Ito prescription for the definition of $dQ$.  $d \neq 1/2$ since we assume that stochastic integration in the definition of $dQ$ is not in Stratonovich way. $\mb{f}^\prime$ is given by Eq.(\ref{fric2}).  Therefore, average heat dissipation rate is changed to
\begin{eqnarray}
h_d(t) &=&\int d\mb{v} \{ -\mb{v} \cdot  \mb{f}^\prime  - d Tr[\mbh{\Gamma} \mbh{\Gamma}^T] \} P(\mb{v},t) \label{hd-1} \\
		&=& \int d\mb{v} \Bigl[ \{ - \frac{1}{2}\partial_j (\Gammah \Gammah^T)_{ij}+\frac{1}{2}(\Gammah \Gammah^T)_{ij}\frac{v_j}{T}-b_i\} v_i  \nonumber \\
	&&- d Tr[\Gammah \Gammah^T] \Bigr] P(t). 
\label{hd-general}
\end{eqnarray}
The average heat dissipation in equilibrium steady state must vanish:
\begin{equation}
h_d(\infty) = \langle \mb{b}\cdot \mb{v}\rangle_e + T(2d-1) \langle Tr[\Xi] \rangle_e=0 
\label{hd-general2}
\end{equation}
The generalized Einstein relation is changed to 
\begin{equation}
\mbox{Eq.(\ref{fric2})} \quad \mbox{and} \quad  \mbox{Eq.(\ref{hd-general2})}.
\end{equation}

Eq.(\ref{entropy-bal2}) still holds without any change.  Substitution of $P_e$ into Eq.(\ref{entropy-bal2}) leads to
\[
\langle \mb{b}\cdot \mb{v} \rangle _e = 0.
\]
Therefore, the generalized Einstein relation can be redefined to 
\begin{equation}
\mbox{Eq.(\ref{fric2})} \quad \mbox{and} \quad \langle \mb{b}\cdot \mb{v} \rangle _e = \langle Tr[\Xi] \rangle_e=0.
\end{equation}
The last equality in the above equation means that the diffusion coefficient of the Fokker-Planck equation, Eq.(\ref{fokker}), becomes traceless on average in an equilibrium steady state.   From Eq.(\ref{hd-1}), both frictional dissipation and heat absorption by fluctuation vanish in the equilibrium steady state since $\langle \mb{v}\cdot \mb{f}^\prime \rangle_e =0$! This is unphysical in dissipative systems. Therefore, Stratonovich prescription is the unique physical choice.

\section{Conclusions and Remarks}
In this manuscript, we have provided quantitative mesoscopic  non-equilibrium thermodynamics of Brownian particles under both multiplicative noise  and feedback control by an external agent.  The dynamics of the Brownian particles is described by a Langevin equation with the multiplicative noise and the feedback control by non-conservative force field.  There is ambiguity in stochastic integration prescription of the Langevin equation due to the multiplicative noise.  However, such ambiguity can be removed by proposing a generalized Einstein relation to guarantee an equilibrium steady state in a closed system: the corresponding Fokker-Planck equation has no  ambiguity.  Statistical properties of the Langevin equation is described unambiguously.  Then, how does one construct thermodynamics without such prescription ambiguity?   Once heat is defined, work and entropy production  are well defined from energy balance and entropy balance.  Thus, we focus on how to define heat dissipated from the particles to their surrounding.  The heat is the energy dissipation of the work done by contact forces between the particles and their surrounding solvent molecules along the position space trajectories of the particles.  There are two ambiguities in the definition of heat.  First, the contact force between the particles and their surrounding depends on the stochastic integration prescription involved in the Langevin equation.  Such ambiguity is removed by the Einstein relation.  Second, the stochastic integration along the trajectories of the particles involved in the calculation of the work done by the contact forces needs to be prescribed.  The Stratonovich prescription  is shown to be  the unique physical choice. 

We remark that  since both Fokker-Planck equation Eq.(\ref{fokker}) and the mesoscopic heat equation Eq.(\ref{stochasticheat})  are expressed in terms of diffusion coefficient $\mbh{\Gamma}\mbh{\Gamma}^T$ and independent to the integration prescription of the Langevin equation, \emph{one can numerically or analytically predict all the thermodynamic quantities by measuring the diffusion coefficient without any mathematical ambiguity}. 

Finally, we gives a comment on the link to fluctuation theorems and Jarzynski equality, which have been studied on Brownian systems with a linear friction, where the thermal noise is not multiplicative.  The proposed thermodynamics in this manuscript shows the possibility to extend the applicability of the fluctuation theorems and Jarzynski equality to the Brownian systems with multiplicative noise \cite{Bochkov81, Rubi87, Crooks99, Crooks00, Hummer01, Seifert05, Lebo, Kurch98, Jarzynski97, Jarzynski97-pre, Jarzynski00, Kim06}.

\begin{acknowledgments}
We thank M.den Nijs and Suk-jin Yoon for useful discussions and comments.  This research is supported by NSF under grant DMR-0341341.
\end{acknowledgments}

\bibliography{nonlin}

\begin{thebibliography}{41}
\expandafter\ifx\csname natexlab\endcsname\relax\def\natexlab#1{#1}\fi
\expandafter\ifx\csname bibnamefont\endcsname\relax
  \def\bibnamefont#1{#1}\fi
\expandafter\ifx\csname bibfnamefont\endcsname\relax
  \def\bibfnamefont#1{#1}\fi
\expandafter\ifx\csname citenamefont\endcsname\relax
  \def\citenamefont#1{#1}\fi
\expandafter\ifx\csname url\endcsname\relax
  \def\url#1{\texttt{#1}}\fi
\expandafter\ifx\csname urlprefix\endcsname\relax\def\urlprefix{URL }\fi
\providecommand{\bibinfo}[2]{#2}
\providecommand{\eprint}[2][]{\url{#2}}

\bibitem[{\citenamefont{Liang et~al.}(2000)\citenamefont{Liang, Medich,
  Czajkowsky, Sheng, Yuan, and Shao}}]{Liang}
\bibinfo{author}{\bibfnamefont{S.}~\bibnamefont{Liang}},
  \bibinfo{author}{\bibfnamefont{D.}~\bibnamefont{Medich}},
  \bibinfo{author}{\bibfnamefont{D.~M.} \bibnamefont{Czajkowsky}},
  \bibinfo{author}{\bibfnamefont{S.}~\bibnamefont{Sheng}},
  \bibinfo{author}{\bibfnamefont{J.}~\bibnamefont{Yuan}}, \bibnamefont{and}
  \bibinfo{author}{\bibfnamefont{Z.}~\bibnamefont{Shao}},
  \bibinfo{journal}{Ultramicroscopy} \textbf{\bibinfo{volume}{84}},
  \bibinfo{pages}{119} (\bibinfo{year}{2000}).

\bibitem[{\citenamefont{Tamayo et~al.}(2001)\citenamefont{Tamayo, Humphris,
  Owen, and Miles}}]{Tamayo}
\bibinfo{author}{\bibfnamefont{J.}~\bibnamefont{Tamayo}},
  \bibinfo{author}{\bibfnamefont{A.~D.~L.} \bibnamefont{Humphris}},
  \bibinfo{author}{\bibfnamefont{R.~J.} \bibnamefont{Owen}}, \bibnamefont{and}
  \bibinfo{author}{\bibfnamefont{M.~J.} \bibnamefont{Miles}},
  \bibinfo{journal}{Biophys. J.} \textbf{\bibinfo{volume}{81}},
  \bibinfo{pages}{526} (\bibinfo{year}{2001}).

\bibitem[{\citenamefont{Kim and Qian}(2004)}]{Kim03}
\bibinfo{author}{\bibfnamefont{K.~H.} \bibnamefont{Kim}} \bibnamefont{and}
  \bibinfo{author}{\bibfnamefont{H.}~\bibnamefont{Qian}},
  \bibinfo{journal}{Phys. Rev. Lett.} \textbf{\bibinfo{volume}{93}},
  \bibinfo{pages}{120602} (\bibinfo{year}{2004}).

\bibitem[{\citenamefont{J{\"u}licher et~al.}(1997)\citenamefont{J{\"u}licher,
  Ajdari, and Prost}}]{Juli97}
\bibinfo{author}{\bibfnamefont{F.}~\bibnamefont{J{\"u}licher}},
  \bibinfo{author}{\bibfnamefont{A.}~\bibnamefont{Ajdari}}, \bibnamefont{and}
  \bibinfo{author}{\bibfnamefont{J.}~\bibnamefont{Prost}},
  \bibinfo{journal}{Rev. Mod. Phys.} \textbf{\bibinfo{volume}{69}},
  \bibinfo{pages}{1269} (\bibinfo{year}{1997}).

\bibitem[{\citenamefont{Colquhoun and Hawkes}(1981)}]{Colqu}
\bibinfo{author}{\bibfnamefont{D.}~\bibnamefont{Colquhoun}} \bibnamefont{and}
  \bibinfo{author}{\bibfnamefont{A.~G.} \bibnamefont{Hawkes}},
  \bibinfo{journal}{Proc. R. Soc. London, Ser. B}
  \textbf{\bibinfo{volume}{211}}, \bibinfo{pages}{205} (\bibinfo{year}{1981}).

\bibitem[{\citenamefont{Urbakh et~al.}(2004)\citenamefont{Urbakh, Klafter,
  Gourdon, and Israelachvili}}]{Urbakh04}
\bibinfo{author}{\bibfnamefont{M.}~\bibnamefont{Urbakh}},
  \bibinfo{author}{\bibfnamefont{J.}~\bibnamefont{Klafter}},
  \bibinfo{author}{\bibfnamefont{D.}~\bibnamefont{Gourdon}}, \bibnamefont{and}
  \bibinfo{author}{\bibfnamefont{J.}~\bibnamefont{Israelachvili}},
  \bibinfo{journal}{Nature} \textbf{\bibinfo{volume}{430}},
  \bibinfo{pages}{525} (\bibinfo{year}{2004}).

\bibitem[{\citenamefont{Kim and Qian}(2006)}]{Kim06}
\bibinfo{author}{\bibfnamefont{K.~H.} \bibnamefont{Kim}} \bibnamefont{and}
  \bibinfo{author}{\bibfnamefont{H.}~\bibnamefont{Qian}}, p.
  \bibinfo{pages}{physics/0601085} (\bibinfo{year}{2006}).

\bibitem[{\citenamefont{Sekimoto}(1997)}]{Seki97}
\bibinfo{author}{\bibfnamefont{K.}~\bibnamefont{Sekimoto}},
  \bibinfo{journal}{J. Phys. Soc. Jpn.} \textbf{\bibinfo{volume}{66}},
  \bibinfo{pages}{1234} (\bibinfo{year}{1997}).

\bibitem[{\citenamefont{Seifert}(2005)}]{Seifert05}
\bibinfo{author}{\bibfnamefont{U.}~\bibnamefont{Seifert}},
  \bibinfo{journal}{Phys. Rev. Lett.} \textbf{\bibinfo{volume}{95}},
  \bibinfo{pages}{040602} (\bibinfo{year}{2005}).

\bibitem[{\citenamefont{Doi and Edwards}(1988)}]{Doi}
\bibinfo{author}{\bibfnamefont{M.}~\bibnamefont{Doi}} \bibnamefont{and}
  \bibinfo{author}{\bibfnamefont{S.~F.} \bibnamefont{Edwards}},
  \emph{\bibinfo{title}{The Theory of Polymer Dynamics}}
  (\bibinfo{publisher}{Oxford University Press, New York},
  \bibinfo{year}{1988}).

\bibitem[{\citenamefont{Hoogerbrugge and Koleman}(1992)}]{Hooger92}
\bibinfo{author}{\bibfnamefont{P.~J.} \bibnamefont{Hoogerbrugge}}
  \bibnamefont{and} \bibinfo{author}{\bibfnamefont{J.~M. V.~A.}
  \bibnamefont{Koleman}}, \bibinfo{journal}{Europhys. Lett.}
  \textbf{\bibinfo{volume}{19}}, \bibinfo{pages}{155} (\bibinfo{year}{1992}).

\bibitem[{\citenamefont{Sancho et~al.}(1982)\citenamefont{Sancho, Miguel, Katz,
  and Gunton}}]{Sancho82}
\bibinfo{author}{\bibfnamefont{J.~M.} \bibnamefont{Sancho}},
  \bibinfo{author}{\bibfnamefont{M.~S.} \bibnamefont{Miguel}},
  \bibinfo{author}{\bibfnamefont{S.~L.} \bibnamefont{Katz}}, \bibnamefont{and}
  \bibinfo{author}{\bibfnamefont{J.~D.} \bibnamefont{Gunton}},
  \bibinfo{journal}{Phys. Rev. A} \textbf{\bibinfo{volume}{26}},
  \bibinfo{pages}{1589} (\bibinfo{year}{1982}).

\bibitem[{\citenamefont{Bier and Astumian}(1996)}]{Bier96}
\bibinfo{author}{\bibfnamefont{M.}~\bibnamefont{Bier}} \bibnamefont{and}
  \bibinfo{author}{\bibfnamefont{R.~D.} \bibnamefont{Astumian}},
  \bibinfo{journal}{Phys. Rev. Lett.} \textbf{\bibinfo{volume}{76}},
  \bibinfo{pages}{4277} (\bibinfo{year}{1996}).

\bibitem[{\citenamefont{Hasty et~al.}(2000)\citenamefont{Hasty, Pradines,
  Dolnik, and Collins}}]{Hasty00}
\bibinfo{author}{\bibfnamefont{J.}~\bibnamefont{Hasty}},
  \bibinfo{author}{\bibfnamefont{J.}~\bibnamefont{Pradines}},
  \bibinfo{author}{\bibfnamefont{M.}~\bibnamefont{Dolnik}}, \bibnamefont{and}
  \bibinfo{author}{\bibfnamefont{J.~J.} \bibnamefont{Collins}},
  \bibinfo{journal}{Proc. Natl. Acad. Sci. USA} \textbf{\bibinfo{volume}{97}},
  \bibinfo{pages}{2075} (\bibinfo{year}{2000}).

\bibitem[{\citenamefont{Qian et~al.}(2002)\citenamefont{Qian, Qian, and
  Tang}}]{Qian-time}
\bibinfo{author}{\bibfnamefont{H.}~\bibnamefont{Qian}},
  \bibinfo{author}{\bibfnamefont{M.}~\bibnamefont{Qian}}, \bibnamefont{and}
  \bibinfo{author}{\bibfnamefont{X.}~\bibnamefont{Tang}}, \bibinfo{journal}{J.
  Stat. Phys.} \textbf{\bibinfo{volume}{107}}, \bibinfo{pages}{1129}
  (\bibinfo{year}{2002}).

\bibitem[{\citenamefont{Arnold}(2000)}]{Arnold00}
\bibinfo{author}{\bibfnamefont{P.}~\bibnamefont{Arnold}},
  \bibinfo{journal}{Phys. Rev. E} \textbf{\bibinfo{volume}{61}},
  \bibinfo{pages}{6091} (\bibinfo{year}{2000}).

\bibitem[{\citenamefont{Shea and Oppenheim}(1996)}]{Shea96}
\bibinfo{author}{\bibfnamefont{J.}~\bibnamefont{Shea}} \bibnamefont{and}
  \bibinfo{author}{\bibfnamefont{I.}~\bibnamefont{Oppenheim}},
  \bibinfo{journal}{J. Phys. Chem.} \textbf{\bibinfo{volume}{100}},
  \bibinfo{pages}{19035} (\bibinfo{year}{1996}).

\bibitem[{\citenamefont{Shea and Oppenheim}(1998)}]{Shea98}
\bibinfo{author}{\bibfnamefont{J.}~\bibnamefont{Shea}} \bibnamefont{and}
  \bibinfo{author}{\bibfnamefont{I.}~\bibnamefont{Oppenheim}},
  \bibinfo{journal}{Physica A} \textbf{\bibinfo{volume}{250}},
  \bibinfo{pages}{265} (\bibinfo{year}{1998}).

\bibitem[{\citenamefont{Hermans}(1981)}]{Hermans}
\bibinfo{author}{\bibfnamefont{J.~J.} \bibnamefont{Hermans}},
  \bibinfo{journal}{Physica A} \textbf{\bibinfo{volume}{109}},
  \bibinfo{pages}{293} (\bibinfo{year}{1981}).

\bibitem[{\citenamefont{Gardiner}(1983)}]{Gardi}
\bibinfo{author}{\bibfnamefont{C.~W.} \bibnamefont{Gardiner}},
  \emph{\bibinfo{title}{Handbook of Stochastic Methods for Physics, Chemistry,
  and the Natural Sciences}} (\bibinfo{publisher}{Springer-Verlag, Berlin},
  \bibinfo{year}{1983}).

\bibitem[{Kim({\natexlab{a}})}]{Kim-prescription}
\bibinfo{note}{$ dQ =-dH = - \nabla_y H \circ dv= - v \circ dv =-v \cdot dv -
  \frac{1}{2} dv \cdot \nabla_v v \cdot dv = -v \cdot (f^\prime dt + \Gamma
  \cdot dB)- \frac{1}{2m}(f^\prime dt + \Gamma \cdot dB) \cdot (f^\prime dt +
  \Gamma \cdot dB)= [-v \cdot f^\prime - \frac{1}{2m} Tr( \Gamma \Gamma^T) ] dt
  - v \cdot \Gamma \cdot dB$, where the second order in $dv$ is kept since
  there is a contribution of the order of $dt$, and $dB_i(t) dB_j(t^\prime) =
  dt \delta_{ij} \delta(t-t^\prime)$ is used at the last step.}

\bibitem[{\citenamefont{Ao and Zhu}(1999)}]{Ao99}
\bibinfo{author}{\bibfnamefont{P.}~\bibnamefont{Ao}} \bibnamefont{and}
  \bibinfo{author}{\bibfnamefont{X.~M.} \bibnamefont{Zhu}},
  \bibinfo{journal}{Phys. Rev. B} \textbf{\bibinfo{volume}{60}},
  \bibinfo{pages}{6850} (\bibinfo{year}{1999}).

\bibitem[{\citenamefont{Klimontovich}(1990)}]{Klimon90}
\bibinfo{author}{\bibfnamefont{Y.~L.} \bibnamefont{Klimontovich}},
  \bibinfo{journal}{Physica A} \textbf{\bibinfo{volume}{163}},
  \bibinfo{pages}{515} (\bibinfo{year}{1990}).

\bibitem[{\citenamefont{Schnakenberg}(1976)}]{Schnak76}
\bibinfo{author}{\bibfnamefont{J.}~\bibnamefont{Schnakenberg}},
  \bibinfo{journal}{Reviews of Modern physics} \textbf{\bibinfo{volume}{48}},
  \bibinfo{pages}{571} (\bibinfo{year}{1976}).

\bibitem[{\citenamefont{Jou et~al.}(1999)\citenamefont{Jou, Casas-V{\'a}zquez,
  and Lebon}}]{Jou99}
\bibinfo{author}{\bibfnamefont{D.}~\bibnamefont{Jou}},
  \bibinfo{author}{\bibfnamefont{J.}~\bibnamefont{Casas-V{\'a}zquez}},
  \bibnamefont{and} \bibinfo{author}{\bibfnamefont{G.}~\bibnamefont{Lebon}},
  \bibinfo{journal}{Rep. Prog. Phys.} \textbf{\bibinfo{volume}{62}},
  \bibinfo{pages}{1035} (\bibinfo{year}{1999}).

\bibitem[{\citenamefont{de~Groot and Mazur}(1984)}]{Mazur}
\bibinfo{author}{\bibfnamefont{S.~R.} \bibnamefont{de~Groot}} \bibnamefont{and}
  \bibinfo{author}{\bibfnamefont{P.}~\bibnamefont{Mazur}},
  \emph{\bibinfo{title}{Non-equilibrium Thermodynamics}}
  (\bibinfo{publisher}{Dover Publications, Inc., New York},
  \bibinfo{year}{1984}).

\bibitem[{\citenamefont{Tolman}(1924)}]{Tolman}
\bibinfo{author}{\bibfnamefont{R.~C.} \bibnamefont{Tolman}},
  \bibinfo{journal}{Phys. Rev.} \textbf{\bibinfo{volume}{23}},
  \bibinfo{pages}{693} (\bibinfo{year}{1924}).

\bibitem[{\citenamefont{Kampen}(1992)}]{Kampen}
\bibinfo{author}{\bibfnamefont{N.~G.~V.} \bibnamefont{Kampen}},
  \emph{\bibinfo{title}{Stochastic Processes in Physics and Chemistry}}
  (\bibinfo{publisher}{North-Holland, Amsterdam}, \bibinfo{year}{1992}).

\bibitem[{\citenamefont{Graham and Haken}(1971)}]{Graham71}
\bibinfo{author}{\bibfnamefont{R.}~\bibnamefont{Graham}} \bibnamefont{and}
  \bibinfo{author}{\bibfnamefont{H.}~\bibnamefont{Haken}}, \bibinfo{journal}{Z.
  Phys. A} \textbf{\bibinfo{volume}{243}}, \bibinfo{pages}{289}
  (\bibinfo{year}{1971}).

\bibitem[{\citenamefont{Ito}(1978)}]{Ito78}
\bibinfo{author}{\bibfnamefont{H.}~\bibnamefont{Ito}}, \bibinfo{journal}{Prog.
  Theor. Phys.} \textbf{\bibinfo{volume}{59}}, \bibinfo{pages}{725}
  (\bibinfo{year}{1978}).

\bibitem[{Kim({\natexlab{b}})}]{Kim-vdep}
\bibinfo{note}{A feedback control on the velocities of macromolecules makes an
  external agent manipulating the control be treated as a Maxwell demon
  \cite{Kim03}. In this case, total entropy change in both the macromolecules
  and their surrounding heat bath can be negative by entropy pumping.}

\bibitem[{\citenamefont{Bochkov and Kuzovlev}(1981)}]{Bochkov81}
\bibinfo{author}{\bibfnamefont{G.~N.} \bibnamefont{Bochkov}} \bibnamefont{and}
  \bibinfo{author}{\bibfnamefont{Y.~E.} \bibnamefont{Kuzovlev}},
  \bibinfo{journal}{Phys. A} \textbf{\bibinfo{volume}{106}},
  \bibinfo{pages}{443} (\bibinfo{year}{1981}).

\bibitem[{\citenamefont{Dufty and Rubi}(1987)}]{Rubi87}
\bibinfo{author}{\bibfnamefont{J.~W.} \bibnamefont{Dufty}} \bibnamefont{and}
  \bibinfo{author}{\bibfnamefont{J.~M.} \bibnamefont{Rubi}},
  \bibinfo{journal}{Phys. Rev. A} \textbf{\bibinfo{volume}{36}},
  \bibinfo{pages}{222} (\bibinfo{year}{1987}).

\bibitem[{\citenamefont{Crooks}(1999)}]{Crooks99}
\bibinfo{author}{\bibfnamefont{G.~E.} \bibnamefont{Crooks}},
  \bibinfo{journal}{Phys. Rev. E} \textbf{\bibinfo{volume}{60}},
  \bibinfo{pages}{2721} (\bibinfo{year}{1999}).

\bibitem[{\citenamefont{Crooks}(2000)}]{Crooks00}
\bibinfo{author}{\bibfnamefont{G.~E.} \bibnamefont{Crooks}},
  \bibinfo{journal}{Phys. Rev. E} \textbf{\bibinfo{volume}{61}},
  \bibinfo{pages}{2361} (\bibinfo{year}{2000}).

\bibitem[{\citenamefont{Hummer and Szabo}(2001)}]{Hummer01}
\bibinfo{author}{\bibfnamefont{G.}~\bibnamefont{Hummer}} \bibnamefont{and}
  \bibinfo{author}{\bibfnamefont{A.}~\bibnamefont{Szabo}},
  \bibinfo{journal}{Proc. Natl. Acad. Sci. USA} \textbf{\bibinfo{volume}{98}},
  \bibinfo{pages}{3658} (\bibinfo{year}{2001}).

\bibitem[{\citenamefont{Lebowitz and Spohn}(1999)}]{Lebo}
\bibinfo{author}{\bibfnamefont{J.~L.} \bibnamefont{Lebowitz}} \bibnamefont{and}
  \bibinfo{author}{\bibfnamefont{H.}~\bibnamefont{Spohn}}, \bibinfo{journal}{J.
  Stat. Phys.} \textbf{\bibinfo{volume}{95}}, \bibinfo{pages}{333}
  (\bibinfo{year}{1999}).

\bibitem[{\citenamefont{Kurchan}(1998)}]{Kurch98}
\bibinfo{author}{\bibfnamefont{J.}~\bibnamefont{Kurchan}}, \bibinfo{journal}{J.
  Phys. A} \textbf{\bibinfo{volume}{31}}, \bibinfo{pages}{3719}
  (\bibinfo{year}{1998}).

\bibitem[{\citenamefont{Jarzynski}(1997{\natexlab{a}})}]{Jarzynski97}
\bibinfo{author}{\bibfnamefont{C.}~\bibnamefont{Jarzynski}},
  \bibinfo{journal}{Phys. Rev. Lett.} \textbf{\bibinfo{volume}{78}},
  \bibinfo{pages}{2690} (\bibinfo{year}{1997}{\natexlab{a}}).

\bibitem[{\citenamefont{Jarzynski}(1997{\natexlab{b}})}]{Jarzynski97-pre}
\bibinfo{author}{\bibfnamefont{C.}~\bibnamefont{Jarzynski}},
  \bibinfo{journal}{Phys. Rev. E} \textbf{\bibinfo{volume}{56}},
  \bibinfo{pages}{5018} (\bibinfo{year}{1997}{\natexlab{b}}).

\bibitem[{\citenamefont{Jarzynski}(2000)}]{Jarzynski00}
\bibinfo{author}{\bibfnamefont{C.}~\bibnamefont{Jarzynski}},
  \bibinfo{journal}{J. Stat. Phys.} \textbf{\bibinfo{volume}{98}},
  \bibinfo{pages}{77} (\bibinfo{year}{2000}).

\end{thebibliography}
\end{document}